\documentclass[aps,prl,twocolumn,superscriptaddress]{revtex4-2}
\usepackage{amsmath,amssymb,amsbsy,graphicx,xcolor,times,subfigure,marvosym,color,bm,multirow,epstopdf}
\usepackage[latin1]{inputenc}
\begin{document}

\title{Geometry dependence of the thermal Hall effect in chiral spin liquids}

\author{G\'abor B. Hal\'asz}
\thanks{This manuscript has been authored by UT-Battelle, LLC, under contract DE-AC05-00OR22725 with the US Department of Energy (DOE). The publisher acknowledges the US government license to provide public access under the DOE Public Access Plan (http://energy.gov/downloads/doe-public-access-plan).}
\affiliation{Materials Science and Technology Division, Oak Ridge National Laboratory, Oak Ridge, TN 37831, USA}
\affiliation{Quantum Science Center, Oak Ridge, TN 37831, USA}


\begin{abstract}

Recent thermal-transport experiments on the Kitaev magnet $\alpha$-RuCl$_3$ highlight the challenge in identifying chiral quantum spin liquids through their quantized thermal Hall effect. Here, we propose that variations in the underlying sample geometry---for example, the introduction of appropriate constrictions---reveal unique aspects of the thermal Hall effect and can be used to determine its origin. By studying standard phenomenological heat-transport equations based on minimal assumptions, we show that, whereas a conventional thermal Hall effect due to, e.g., phonons or magnons is completely geometry independent, a thermal Hall effect originating from a chiral fermion edge mode is significantly enhanced by constrictions at low temperatures. This unique geometry-dependent signature provides a practical approach for identifying chiral spin liquids in candidate materials like $\alpha$-RuCl$_3$ using currently available thermal-transport experiments.

\end{abstract}


\maketitle


\emph{Introduction.}---Quantum spin liquids are highly entangled magnetic insulators hosting fractionalized anyonic quasiparticles coupled to emergent gauge fields~\cite{Balents-2010,Savary-2016,Zhou-2017,Knolle-2019,Broholm-2020}. Because of their fundamentally nonlocal nature, these anyonic quasiparticles in quantum spin liquids are promising from the perspective of topological quantum computation~\cite{Kitaev-2003,Nayak-2008,Klocke-2024}. For the same reason, however, they are extremely difficult to detect in real materials as local spectroscopic probes like neutron scattering always excite them in pairs and observe a diffuse continuum response whose interpretation requires detailed modelling~\cite{Scheie-2024}.

In this regard, non-time-reversal-symmetric ``chiral'' quantum spin liquids---like the Kalmeyer-Laughlin state~\cite{Kalmeyer-1987,Kalmeyer-1989} or the non-Abelian Kitaev spin liquid~\cite{Kitaev-2006}---are amenable to more direct experimental identification as they support chiral fermion edge modes in addition to the intimately related bulk anyons. These chiral edge modes allow for unidirectional energy transport along the boundary and give rise to a quantized thermal Hall effect~\cite{Guo-2022,Zhang-2024}---an analog of the quantized electrical Hall effect in quantum Hall states---that provides a universal hallmark for the given chiral spin liquid.

However, as exemplified by the recent controversy around thermal-transport experiments in the honeycomb Kitaev magnet $\alpha$-RuCl$_3$~\cite{Kasahara-2018,Yamashita-2020,Czajka-2021,Yokoi-2021,Tanaka-2022,Bruin-2022a,Lefrancois-2022,Bruin-2022b,Czajka-2023,Lefrancois-2023,Zhang-2023,Zhang-2024a,Imamura-2024,Zhang-2024b}, identifying a chiral spin liquid by its quantized thermal Hall effect is far from straightforward in a real material. Although a half-integer-quantized thermal Hall effect---consistent with the non-Abelian Kitaev spin liquid---was observed in the initial experiments~\cite{Kasahara-2018,Yamashita-2020,Yokoi-2021}, other groups failed to reproduce this quantization and instead attributed the thermal Hall effect to topological magnons~\cite{Czajka-2023} or a conventional phonon mechanism~\cite{Lefrancois-2022}. At the same time, it was theoretically pointed out that a finite heat exchange between a chiral fermion edge mode and bulk phonons---which further complicates the interpretation---is not only unavoidable but in fact necessary for observing the quantization of the thermal Hall effect~\cite{Ye-2018,Vinkler-Aviv-2018}. To disentangle the quantized edge contribution from the diffusive phonon transport, Ref.~\onlinecite{Klocke-2022} next proposed mixed mesoscopic-macroscopic setups for thermal-transport measurements; see also Refs.~\onlinecite{Klocke-2021,Wei-2021,Wei-2023} for closely related schemes. Due to the enormous mismatch in length scales between the mesoscopic and macroscopic regions, however, the fabrication of these unconventional thermal-transport setups poses a major experimental challenge.

\begin{figure}[b]
\includegraphics[width=0.92\columnwidth]{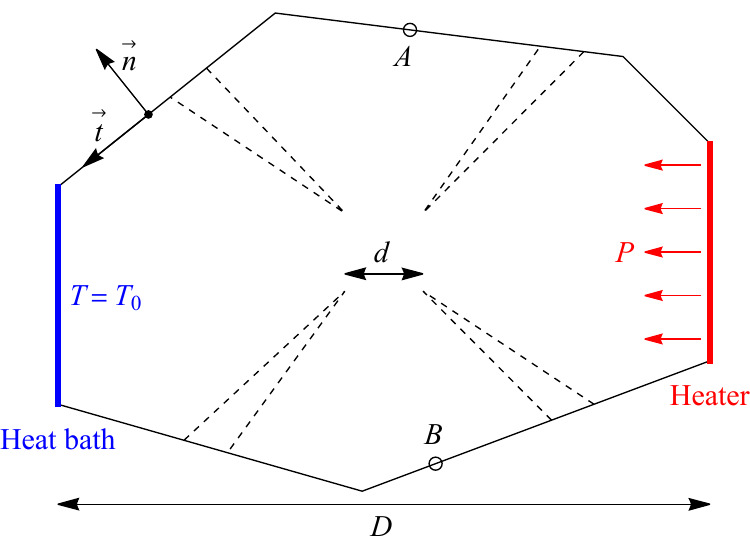}
\caption{Thermal-transport setup in a general two-dimensional geometry where the sample of size $D$ can be constricted by cutting wedges along the dashed lines to create four quadrants connected by a central region of size $d < D$. With a heater applying power $P$ to one end of the sample and a heat bath keeping the other end at a fixed temperature $T_0$, the thermal Hall resistance is measured via the transverse temperature difference between points $A$ and $B$. The normal ($\vec{n}$) and tangential ($\vec{t}$) unit vectors along a section of the edge are also shown.} \label{fig-1}
\end{figure}

In this Letter, we propose to distinguish between different origins of the thermal Hall effect in insulating magnets~\cite{Guo-2022} by varying the geometry in a fully macroscopic setup through, for example, creating constrictions, as depicted in Fig.~\ref{fig-1}. By considering standard phenomenological heat-transport equations based on minimal assumptions, we show that, for any traditional ``bulk'' thermal Hall effect---from phonons or magnons, due to an intrinsic or extrinsic mechanism---the experimentally measured thermal Hall resistance is not affected at all by such a change in the geometry. In contrast, for a thermal Hall effect originating from a chiral fermion edge mode in a chiral spin liquid, the lack of thermalization between the edge mode and the bulk phonons at low temperatures~\cite{Ye-2018} gives rise to a marked increase of the thermal Hall resistance upon creating constrictions in the sample. Hence, as the temperature is lowered, the thermal Hall resistance in these constricted samples is not suppressed straight away---as observed in all previous experiments on unconstricted samples of $\alpha$-RuCl$_3$~\cite{Kasahara-2018,Yamashita-2020,Czajka-2021,Yokoi-2021,Tanaka-2022,Bruin-2022a,Lefrancois-2022,Bruin-2022b,Czajka-2023,Lefrancois-2023,Zhang-2023,Zhang-2024a,Imamura-2024,Zhang-2024b}--- but instead features a pronounced peak before eventually being suppressed. This qualitative difference in the thermal Hall effect between constricted and unconstricted samples is a distinctive signature of chiral fermion edge modes and provides a viable approach for experimentally validating chiral spin liquids in candidate materials using currently available thermal-transport measurements; see an experimental companion paper~\cite{Zhang-2025} for a concrete demonstration in $\alpha$-RuCl$_3$.

\emph{Thermal Hall effect.}---We consider thermal transport in a general two-dimensional geometry, as shown in Fig.~\ref{fig-1}, where a heater applies power $P$ to one end of the sample, while a heat bath keeps the other end at a fixed temperature $T_0$. Assuming a small uniform sample thickness in the third (lateral) dimension, we use a simplified two-dimensional description of the heat transport where any lateral dependence is neglected and the power $P$ is defined per unit thickness. We further assume that the heat transport is dominated by a longitudinal component due to diffusive phonons and that spatial variations in the temperature $T$ across the sample are much smaller than the base temperature $T_0$. For a traditional bulk thermal Hall effect corresponding to, e.g., phonons~\cite{Strohm-2005,Li-2020,Grissonnanche-2020,Boulanger-2020} or magnons~\cite{Onose-2010,Hirschberger-2015}, the bulk heat-current density, $\vec{j} = (j_x, j_y)$, can then be written in terms of the temperature gradient $\vec{\nabla} T = (\partial_x T, \partial_y T)$ as
\begin{equation}
\vec{j} = -\begin{pmatrix} \kappa_{xx} & \kappa_{xy} \\ -\kappa_{xy} & \kappa_{xx} \end{pmatrix} \cdot \vec{\nabla} T, \label{eq-kappa}
\end{equation}
where $\kappa_{xx}$ is the longitudinal thermal conductivity, and $\kappa_{xy}$ is the thermal Hall conductivity. Assuming a steady state, energy conservation in the bulk implies $\vec{\nabla} \cdot \vec{j} = -\kappa_{xx} \nabla^2 T = 0$, while energy conservation along a free section of the edge---with no heater or heat bath connected to it---means that the normal component of $\vec{j}$ vanishes. For a heated section of the edge, this normal component, $\vec{n} \cdot \vec{j}$, is instead given by minus the applied power per unit length, $p$, whose line integral along the entire heated section is the power $P$. Since Eq.~(\ref{eq-kappa}) gives
\begin{equation}
\vec{n} \cdot \vec{j} = -\kappa_{xx} \, \vec{n} \cdot \vec{\nabla} T - \kappa_{xy} \, \vec{t} \cdot \vec{\nabla} T, \label{eq-n}
\end{equation}
the position dependence of the temperature $T$ inside the sample is determined by the following set of differential equations and boundary conditions:
\begin{align}
\nabla^2 T &= 0 \qquad (\textrm{Bulk}), \nonumber \\
\kappa_{xx} \, \vec{n} \cdot \vec{\nabla} T + \kappa_{xy} \, \vec{t} \cdot \vec{\nabla} T &= \bigg\{ \begin{matrix} 0 \quad \, (\textrm{Free edge}), \quad \, \\ p \quad (\textrm{Heated edge}), \end{matrix} \label{eq-T-1} \\
T &= T_0 \quad \,\, (\textrm{Heat-bath edge}), \nonumber
\end{align}
where $\vec{n}$ and $\vec{t}$ are the normal and tangential unit vectors along the given section of the edge (see Fig.~\ref{fig-1}).

The existence of a finite thermal Hall conductivity $\kappa_{xy}$ requires time-reversal symmetry to be broken by, e.g., an external magnetic field, which means that $\kappa_{xy}$ changes sign if the field is reversed. The thermal Hall resistance---the quantity directly measured in thermal-transport experiments---is the ratio of the field-antisymmetrized transverse temperature difference to the total heat current (i.e., the applied power),
\begin{equation}
\Lambda_{AB} = \frac{1} {2P} \Big\{ \left[ T_A^{+} - T_B^{+} \right] - \left[ T_A^{-} - T_B^{-} \right] \Big\}, \label{eq-Lambda-1}
\end{equation}
where the temperatures $T_{A,B}^{+}$ at points $A$ and $B$ along the edge (see Fig.~\ref{fig-1}) are found by solving Eq.~(\ref{eq-T-1}) directly, while the analogous temperatures $T_{A,B}^{-}$ are obtained from Eq.~(\ref{eq-T-1}) after the substitution $\kappa_{xy} \to -\kappa_{xy}$. Note that the points $A$ and $B$ belong to the two opposite sections of the edge, as defined by the locations of the heater and the heat bath.

\emph{Chiral spin liquids.}---In chiral quantum spin liquids like the non-Abelian Kitaev spin liquid~\cite{Kitaev-2006}, there is another kind of thermal Hall effect that originates from a chiral fermion edge mode and does not generically conform to the purely bulk description above. As discussed in Ref.~\onlinecite{Ye-2018}, the heat exchange between the bulk phonons and the chiral fermion edge mode is strongly suppressed at low temperatures, which necessitates the introduction of distinct temperatures for the bulk phonons ($T_{\mathrm{bulk}}$) and the fermion edge mode ($T_{\mathrm{edge}}$). In the absence of a traditional bulk thermal Hall effect, the bulk heat-current density carried by the phonons is then simply given by $\vec{j}_{\mathrm{bulk}} = -\kappa_{xx} \vec{\nabla} T_{\mathrm{bulk}}$. At the same time, there is also an edge heat current $I_{\mathrm{edge}}$ (defined per unit lateral thickness) that is carried by the chiral fermion edge mode in the tangential direction $\vec{t}$ (see Fig.~\ref{fig-1}) all around the boundary. This edge heat current depends on the edge temperature and can be linearized around the base temperature $T_0$ as~\cite{Ye-2018}
\begin{equation}
I_{\mathrm{edge}} (T_{\mathrm{edge}}) = I_{\mathrm{edge}} (T_0) + \kappa_{\mathrm{edge}} (T_0) [T_{\mathrm{edge}} - T_0], \label{eq-I}
\end{equation}
where $\kappa_{\mathrm{edge}} (T_0) = a^{-1} \kappa_{\mathrm{quant}} (T_0)$ in terms of the thickness $a$ of a single spin-liquid layer and its quantized thermal Hall conductance, $\kappa_{\mathrm{quant}} (T_0) = \pi c k_B^2 T_0 / 6 \hbar$, with the chiral central charge $c$. Importantly, the edge heat current changes along the tangential direction, $\vec{t} \cdot \vec{\nabla} I_{\mathrm{edge}} \neq 0$, if heat is exchanged between the bulk phonons and the fermion edge mode due to a local temperature mismatch between the two degrees of freedom: $T_{\mathrm{bulk}} \neq T_{\mathrm{edge}}$. By linearizing this heat exchange in the temperature mismatch with a general temperature-dependent coefficient $\lambda (T_0)$, the edge temperature is then found to be decaying to the bulk temperature,
\begin{equation}
\vec{t} \cdot \vec{\nabla} T_{\mathrm{edge}} = -\ell^{-1} \left( T_{\mathrm{edge}} - T_{\mathrm{bulk}} \right), \label{eq-T-2}
\end{equation}
on a temperature-dependent edge-bulk thermalization length scale $\ell (T_0) = \kappa_{\mathrm{edge}} (T_0) / \lambda (T_0)$ that diverges in the $T_0 \to 0$ limit~\cite{Ye-2018}. Furthermore, energy conservation along a free section of the edge means that any change of the edge heat current in the tangential direction, $\vec{t} \cdot \vec{\nabla} I_{\mathrm{edge}} \approx \kappa_{\mathrm{edge}} \, \vec{t} \cdot \vec{\nabla} T_{\mathrm{edge}}$ [see Eq.~(\ref{eq-I})], must be provided by the normal component of bulk heat-current density, $\vec{n} \cdot \vec{j}_{\mathrm{bulk}} = -\kappa_{xx} \, \vec{n} \cdot \vec{\nabla} T_{\mathrm{bulk}}$, with an additional contribution coming from the applied power per unit length, $p$, along a heated section. Therefore, the generalizations of the differential equations and boundary conditions in Eq.~(\ref{eq-T-1}) to chiral spin liquids with distinct bulk and edge temperatures read
\begin{align}
\nabla^2 T_{\mathrm{bulk}} &= 0 \qquad (\textrm{Bulk}), \label{eq-T-3} \\
\kappa_{xx} \, \vec{n} \cdot \vec{\nabla} T_{\mathrm{bulk}} + \kappa_{\mathrm{edge}} \, \vec{t} \cdot \vec{\nabla} T_{\mathrm{edge}} &= \bigg\{ \begin{matrix} 0 \quad \, (\textrm{Free edge}), \quad \, \\ p \quad (\textrm{Heated edge}), \end{matrix} \nonumber \\
T_{\mathrm{bulk}} &= T_0 \quad \,\, (\textrm{Heat-bath edge}). \nonumber
\end{align}
Note that the heater and the heat bath are assumed to couple to the bulk phonons (rather than the fermion edge mode) which explains that the heater power $p$ does not appear in Eq.~(\ref{eq-T-2}) and that the heat bath fixes $T_{\mathrm{bulk}}$ in Eq.~(\ref{eq-T-3}).

\begin{figure*}[t]
\includegraphics[width=1.9\columnwidth]{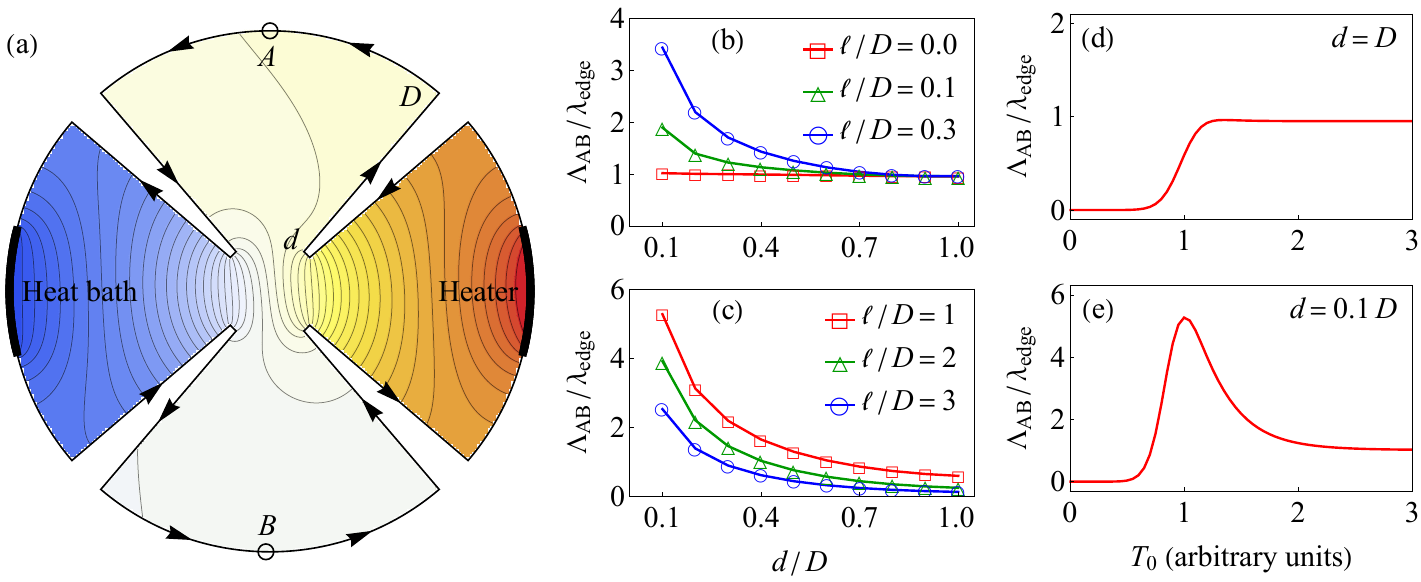}
\caption{(a) Concrete geometry of outer radius $D$ and inner radius $d$ for solving the heat-transport equations of a chiral spin liquid in Eqs.~(\ref{eq-T-2}) and (\ref{eq-T-3}). The heat map shows $T_{\mathrm{bulk}}$ for $d/D = 0.2$, $\ell/D = 1$, and $\kappa_{\mathrm{edge}} = 0.1 \kappa_{xx}$, while the arrows mark the direction of the chiral fermion edge mode. (b-e) Geometry-dependent prefactor $\theta (\ell/D, d/D)$ of the thermal Hall resistance $\Lambda_{AB}$ [see Eq.~(\ref{eq-Lambda-3})] against (b,c) the radius ratio $d/D$ for edge-bulk thermalization lengths (b) $\ell < D$ and (c) $\ell \geq D$, and (d,e) the base temperature $T_0$ for (d) unconstricted and (e) constricted samples. Here we assume $\ell \propto T_0^{-5}$~\cite{Ye-2018} with $T_0 = 1$ corresponding to $\ell = D$.} \label{fig-2}
\end{figure*}

In Ref.~\onlinecite{Ye-2018}, the heat-transport equations in Eqs.~(\ref{eq-T-2}) and (\ref{eq-T-3}) were solved in a standard rectangular geometry. It was found that, if the edge-bulk thermalization length $\ell$ is much smaller than the sample size, the chiral fermion edge mode imprints the bulk phonons with a transverse temperature gradient that could also be explained by a traditional bulk thermal Hall effect [see Eq.~(\ref{eq-kappa})] with $\kappa_{xy} = \kappa_{\mathrm{edge}}$. In this work, we provide a more rigorous basis for this result and generalize it to arbitrary geometry by pointing out that, in the limit of vanishing edge-bulk thermalization length ($\ell \to 0$), the ``edge'' heat-transport equations in Eqs.~(\ref{eq-T-2}) and (\ref{eq-T-3}) directly reduce to the corresponding ``bulk'' heat-transport equations in Eq.~(\ref{eq-T-1}). Indeed, Eq.~(\ref{eq-T-2}) readily implies $T_{\mathrm{bulk}} = T_{\mathrm{edge}}$ in the limit of $\ell \to 0$, while substituting $T_{\mathrm{bulk}} = T_{\mathrm{edge}} \to T$ and $\kappa_{\mathrm{edge}} \to \kappa_{xy}$ in Eq.~(\ref{eq-T-3}) then immediately recovers Eq.~(\ref{eq-T-1}). Remarkably, this result implies that a thermal Hall effect coming from a chiral fermion edge mode is even theoretically indistinguishable from a traditional bulk thermal Hall effect, provided that the fermion edge mode is locally thermalized with the bulk phonons.

\emph{Geometry dependence.}---We aim to understand how the experimentally measured thermal Hall resistance in Eq.~(\ref{eq-Lambda-1}) depends on the sample geometry, which may be modified by, for example, creating constrictions in the sample (see Fig.~\ref{fig-1}). We first show that the thermal Hall resistance for a traditional bulk thermal Hall effect---or, equivalently, an edge thermal Hall effect with local edge-bulk thermalization---is entirely geometry independent. To this end, we start by taking the inverse of Eq.~(\ref{eq-kappa}) and writing it in terms of the lateral unit vector $\vec{z}$ as
\begin{equation}
\vec{\nabla} T = -\begin{pmatrix} \lambda_{xx} & -\lambda_{yx} \\ \lambda_{yx} & \lambda_{xx} \end{pmatrix} \cdot \vec{j} = -\lambda_{xx} \, \vec{j} - \lambda_{yx} \, \vec{z} \times \vec{j}, \label{eq-lambda}
\end{equation}
where $\lambda_{xx}$ is the longitudinal thermal resistivity, and $\lambda_{yx}$ is the thermal Hall resistivity. Noting that $\lambda_{yx}$ changes sign upon reversing the external magnetic field, the transverse temperature differences in Eq.~(\ref{eq-Lambda-1}) are then given by
\begin{equation}
T_A^{\pm} - T_B^{\pm} = -\lambda_{xx} \int_B^A d\vec{r} \cdot \vec{j} \pm \lambda_{yx} \int_B^A \vec{z} \cdot \left( d\vec{r} \times \vec{j} \right). \label{eq-TAB}
\end{equation}
Since the first term cancels between $T_A^{+} - T_B^{+}$ and $T_A^{-} - T_B^{-}$, while the line integral of $\vec{z} \cdot (d\vec{r} \times \vec{j})$ from $B$ to $A$ is simply the total heat current (i.e., the applied power $P$), the experimentally measured thermal Hall resistance in Eq.~(\ref{eq-Lambda-1}) is found to be directly given by the thermal Hall resistivity:
\begin{equation}
\Lambda_{AB} = \frac{1} {2P} \left[ \lambda_{yx} P - (-\lambda_{yx} P) \right] = \lambda_{yx}. \label{eq-Lambda-2}
\end{equation}
Note that these two quantities have the same dimensions because $P$ is defined per unit thickness. Since the thermal Hall resistivity $\lambda_{yx}$ is exclusively a property of the material, the lack of geometry-dependent prefactors (e.g., aspect ratios) in Eq.~(\ref{eq-Lambda-2}) means that the thermal Hall resistance $\Lambda_{AB}$ does not have any geometry dependence. This result has two profound consequences. First, the thermal Hall resistance stays exactly the same if the sample geometry is modified by, e.g., creating constrictions. Second, $\Lambda_{AB}$ is not sensitive to the precise locations of points $A$ and $B$, as long as they are on the two opposite sections of the edge with respect to the heater and the heat bath (see Fig.~\ref{fig-1}). Note that, for constricted samples, it is implicitly assumed that the phonons are still diffusive in the small central region of size $d < D$.

For a chiral spin liquid with a finite edge-bulk thermalization length $\ell$, in contrast, the thermal Hall resistance $\Lambda_{AB}$ acquires a nontrivial geometry dependence. Remembering that the $\ell \to 0$ limit is equivalent to a traditional bulk thermal Hall effect, it is instructive to compare $\Lambda_{AB}$ to an effective thermal Hall resistivity $\lambda_{\mathrm{edge}}$ that corresponds to $\lambda_{yx}$ in Eq.~(\ref{eq-lambda}) and is defined by inverting Eq.~(\ref{eq-kappa}) with $\kappa_{xy} = \kappa_{\mathrm{edge}}$:
\begin{equation}
\begin{pmatrix} \lambda_{xx} & -\lambda_{\mathrm{edge}} \\ \lambda_{\mathrm{edge}} & \lambda_{xx} \end{pmatrix} = \begin{pmatrix} \kappa_{xx} & \kappa_{\mathrm{edge}} \\ -\kappa_{\mathrm{edge}} & \kappa_{xx} \end{pmatrix}^{-1}. \label{eq-lambda-kappa}
\end{equation}
For an unconstricted sample geometry with a single characteristic size $D$, it was found in Ref.~\onlinecite{Ye-2018} that, while $\Lambda_{AB} \approx \lambda_{\mathrm{edge}}$ at $\ell \ll D$ [cf.~Eq.~(\ref{eq-Lambda-2})], the thermal Hall resistance is strongly suppressed, $\Lambda_{AB} \ll \lambda_{\mathrm{edge}}$, if the thermalization length exceeds the sample dimension, $\ell \gg D$. The underlying reason is that, for $\ell \gg D$, the chiral fermion edge mode is thermally decoupled from the bulk phonons, i.e., it cannot effectively heat up (cool down) around the heater (heat bath) or imprint the bulk phonons with any temperature gradient. Importantly, $\Lambda_{AB}$ is obtained by assuming that the thermometers at points $A$ and $B$ measure the temperature of the bulk phonons such that $T_{A,B}^{\pm}$ in Eq.~(\ref{eq-Lambda-1}) correspond to $T_{\mathrm{bulk}}$.

The geometry dependence is even more complex in a constricted geometry where the sample of overall size $D$ is effectively divided into four quadrants connected by a small central region of size $d < D$ (see Fig.~\ref{fig-1}). Specifically, due to the two different characteristic sizes $D$ and $d$, there are three distinct regimes for the thermal Hall resistance $\Lambda_{AB}$; in addition to $\Lambda_{AB} \approx \lambda_{\mathrm{edge}}$ for $\ell \ll d$ and $\Lambda_{AB} \ll \lambda_{\mathrm{edge}}$ for $\ell \gg D$, there is also an intermediate regime with $d < \ell < D$ where we anticipate the thermal Hall resistance to be larger than in either of the other two regimes: $\Lambda_{AB} > \lambda_{\mathrm{edge}}$. In this regime, since $\ell < D$, the chiral fermion edge mode can properly heat up around the heater and cool down around the heat bath. At the same time, since $\ell > d$, it does not thermalize well inside the central region and remains hot (cold) as it enters the ``downstream'' (``upstream'') quadrant containing point $A$ ($B$) after leaving the heater (heat-bath) quadrant. Moreover, because of the constriction, the heat deposited in the downstream quadrant cannot easily escape, and the heat absorbed from the upstream quadrant cannot be easily replaced, which leads to an increase in the transverse temperature difference and, hence, an enhanced thermal Hall resistance.

To confirm our expectations for the geometry dependence, we consider the representative geometry in Fig.~\ref{fig-2}(a) and numerically solve the heat-transport equations in Eqs.~(\ref{eq-T-2}) and (\ref{eq-T-3}) using the boundary-element method described in the Supplemental Material~\cite{SM}. For the experimentally relevant regime, $\kappa_{\mathrm{edge}} \ll \kappa_{xx}$ and $\lambda_{\mathrm{edge}} \approx \kappa_{\mathrm{edge}} / \kappa_{xx}^2 \ll \lambda_{xx}$, the thermal Hall resistance takes the general form [cf.~Eq.~(\ref{eq-Lambda-2})]
\begin{equation}
\Lambda_{AB} = \theta \big( \ell/D, d/D \big) \, \lambda_{\mathrm{edge}}, \label{eq-Lambda-3}
\end{equation}
where the geometry dependence is entirely captured in the dimensionless function $\theta (\ell/D, d/D)$. In Figs.~\ref{fig-2}(b,c), this function is plotted against the constriction size $d$ for various edge-bulk thermalization lengths $\ell$. In the limit of $\ell \to 0$, we observe $\theta \approx 1$ regardless of $d$, which recovers our previous result for the lack of geometry dependence in the presence of local edge-bulk thermalization. For finite $\ell < D$, while $\theta$ remains approximately $1$ at $d > \ell$, geometry dependence sets in through a marked upturn of $\theta$ at $d \lesssim \ell$. For $\ell \gtrsim D$, geometry dependence then extends to the entire range of $d < D$ and becomes progressively stronger with increasing $\ell$.

Finally, Figs.~\ref{fig-2}(d,e) show the same dimensionless quantity, $\Lambda_{AB} / \lambda_{\mathrm{edge}}$, against the base temperature $T_0$ for both unconstricted ($d = D$) and constricted ($d < D$) samples. Here we assume that the edge-bulk thermalization length depends on the base temperature via $\ell \propto T_0^{-5}$~\cite{Ye-2018}. In the unconstricted case [see Fig.~\ref{fig-2}(d)], the thermal Hall resistance $\Lambda_{AB}$---in units of the effective thermal Hall resistivity $\lambda_{\mathrm{edge}} \approx \kappa_{\mathrm{edge}} / \kappa_{xx}^2$---is monotonically suppressed as the temperature is lowered. In the constricted case [see Fig.~\ref{fig-2}(e)], however, the thermal Hall resistance exhibits a pronounced peak before eventually being suppressed at the lowest temperatures.

\emph{Discussion.}---By studying the geometry dependence of the thermal Hall effect in chiral spin liquids---and more generally, insulating magnets---through a purely phenomenological approach, we found a qualitative difference between a thermal Hall effect originating from a chiral fermion edge mode and a traditional bulk thermal Hall effect. While the experimentally measured thermal Hall resistance is geometry independent in the latter case, it shows a dramatic increase upon constricting the sample via appropriately positioned wedges (see Fig.~\ref{fig-1}) in the former case, provided a finite edge-bulk temperature mismatch, i.e., a finite edge-bulk thermalization length $\ell$.

We now discuss the implications of our study for the three potential origins of the thermal Hall effect that were put forward for $\alpha$-RuCl$_3$. The conventional phonon mechanism suggested in Ref.~\onlinecite{Lefrancois-2022} clearly corresponds to a bulk thermal Hall effect and is therefore not expected to show any geometry dependence. The thermal Hall effect from topological magnons, as proposed by Ref.~\onlinecite{Czajka-2023}, was originally introduced as a bulk mechanism~\cite{Katsura-2010} but then later argued to be carried by chiral edge modes along the boundary~\cite{Matsumoto-2011}. Nevertheless, any such finite-energy magnon edge modes are expected to overlap and hybridize with the bulk magnon modes~\cite{Habel-2024}, thus realizing local edge-bulk thermalization, i.e., $\ell \to 0$. In turn, we showed that an edge thermal Hall effect with $\ell \to 0$ is equivalent to a bulk thermal Hall effect and is still not expected to exhibit any geometry dependence. For a zero-energy fermion edge mode, however, the edge-bulk thermalization length $\ell$ diverges at the lowest temperatures~\cite{Ye-2018}. The corresponding thermal Hall effect~\cite{Kasahara-2018,Yamashita-2020,Yokoi-2021} is then expected to show a qualitative dependence on the sample geometry; while the thermal Hall resistance is suppressed monotonically upon lowering the temperature in an unconstrained sample [see Fig.~\ref{fig-2}(d)], it features a pronounced peak before eventually being suppressed in a constrained sample [see Fig.~\ref{fig-2}(e)]. As such, the geometry dependence of the thermal Hall effect can distinguish this interesting scenario from the two competing interpretations, thereby providing validation for a chiral spin liquid in $\alpha$-RuCl$_3$~\cite{Zhang-2025}.

\emph{Acknowledgments.}---We thank Jason Alicea, Leon Balents, Kai Klocke, Joel Moore, Lucile Savary, and Mengxing Ye for collaboration on closely related earlier works. We further acknowledge Michael McGuire, Jiaqiang Yan, and Heda Zhang for inspiring experimental discussions. This material is based upon work supported by the U.S. Department of Energy, Office of Science, National Quantum Information Science Research Centers, Quantum Science Center.



\clearpage

\begin{widetext}

\subsection{\large Supplemental Material}

\section{Boundary-element method for solving the heat-transport equations}

We solve the heat-transport equations in Eqs.~(6) and (7) of the main text using a standard boundary-element method. The key idea behind this approach is to reduce the dimensionality of the problem by transforming a two-dimensional partial differential equation into an appropriate integral equation along the one-dimensional boundary of the system. In the specific case of a two-dimensional Poisson's equation $\nabla^2 T (\vec{r} \,) = 0$ [see Eqs.~(3) and (7) in the main text] defined inside a finite region $V$ with a single boundary $\partial V$, the starting point is Green's second identity,
\begin{equation}
\int_V d^2 r \left[ T (\vec{r} \,) \, \nabla^2 G_{\vec{r}_0} (\vec{r} \,) - G_{\vec{r}_0} (\vec{r} \,) \, \nabla^2 T (\vec{r} \,) \right] = \oint_{\partial V} dr \left[ T (\vec{r} \,) \, \partial_{\vec{n}} G_{\vec{r}_0} (\vec{r} \,) - G_{\vec{r}_0} (\vec{r} \,) \, \partial_{\vec{n}} T (\vec{r} \,) \right], \label{eq-supp-G-1}
\end{equation}
where $ \partial_{\vec{n}}$ is the derivative along the normal direction $\vec{n}$ of the boundary, and $G_{\vec{r}_0} (\vec{r} \,) = (2\pi)^{-1} \ln |\vec{r} - \vec{r}_0|$ is the Green's function of the Poisson's equation corresponding to a given point $\vec{r}_0$ inside the region $V$. Since $\nabla^2 T (\vec{r} \,) = 0$ by the Poisson's equation and $\nabla^2 G_{\vec{r}_0} (\vec{r} \,) = \delta (\vec{r} - \vec{r}_0)$ by construction, Green's second identity in Eq.~(\ref{eq-supp-G-1}) reduces to
\begin{equation}
T (\vec{r}_0) = \oint_{\partial V} dr \left[ T (\vec{r} \,) \, \partial_{\vec{n}} G_{\vec{r}_0} (\vec{r} \,) - G_{\vec{r}_0} (\vec{r} \,) \, \partial_{\vec{n}} T (\vec{r} \,) \right]. \label{eq-supp-G-2}
\end{equation}
If the point $\vec{r}_0$ is then taken to the boundary $\partial V$ from inside the region $V$, the integral equation in Eq.~(\ref{eq-supp-G-2}) is indeed restricted to the boundary, relating $T (\vec{r} \,)$ and $\partial_{\vec{n}} T (\vec{r} \,)$ for various $\vec{r} \in \partial V$.

To numerically solve Eqs.~(6) and (7) in the main text using the boundary-element method, we first discretize the edge into $N$ straight segments $j = 1, \ldots, N$, where each segment $j$ connects two points $\vec{r}_j$ and $\vec{r}_{j+1}$ on the edge with length $d_j = |\vec{r}_{j+1} - \vec{r}_j|$ and central point $\vec{R}_j = (\vec{r}_j + \vec{r}_{j+1}) / 2$. The bulk and edge temperatures as well as the normal derivatives of the bulk temperatures at the central points can each be arranged in an $N$-dimensional vector:
\begin{align}
\mathbf{T}_{\mathrm{bulk}} &= \left[ T_{\mathrm{bulk}} (\vec{R}_1), \ldots, T_{\mathrm{bulk}} (\vec{R}_N) \right], \nonumber \\
\mathbf{T}_{\mathrm{edge}} &= \left[ T_{\mathrm{edge}} (\vec{R}_1), \ldots, T_{\mathrm{edge}} (\vec{R}_N) \right], \label{eq-supp-T-1} \\
\partial \mathbf{T}_{\mathrm{bulk}} &= \left[ \partial_{\vec{n}} T_{\mathrm{bulk}} (\vec{R}_1), \ldots, \partial_{\vec{n}} T_{\mathrm{bulk}} (\vec{R}_N) \right]. \nonumber
\end{align}
If we assume that both $T_{\mathrm{bulk}} (\vec{r} \,)$ and $\partial_{\vec{n}} T_{\mathrm{bulk}} (\vec{r} \,)$ are approximately constant along each segment $j$, the discretized version of Eq.~(\ref{eq-supp-G-2}) corresponding to the bulk Poisson's equation $\nabla^2 T_{\mathrm{bulk}} (\vec{r} \,) = 0$ in Eq.~(7) of the main text becomes
\begin{equation}
T_{\mathrm{bulk}} (\vec{r}_0) = \sum_{j=1}^N \left[ T_{\mathrm{bulk}} (\vec{R}_j) \int_{\vec{r}_j}^{\vec{r}_{j+1}} dr \, \partial_{\vec{n}} G_{\vec{r}_0} (\vec{r} \,) - \partial_{\vec{n}} T_{\mathrm{bulk}} (\vec{R}_j) \int_{\vec{r}_j}^{\vec{r}_{j+1}} dr \, G_{\vec{r}_0} (\vec{r} \,) \right]. \label{eq-supp-G-3}
\end{equation}
If the point $\vec{r}_0$ is then taken to each $\vec{R}_j$ from inside the edge, the resulting equations can be written in the compact vector form,
\begin{equation}
\partial \mathbf{G} \cdot \mathbf{T}_{\mathrm{bulk}} - \mathbf{G} \cdot \partial \mathbf{T}_{\mathrm{bulk}} = 0, \label{eq-supp-T-2}
\end{equation}
where the elements of the $N \times N$ matrices $\mathbf{G}$ and $\partial \mathbf{G}$ are approximately given by
\begin{align}
G_{j,k} &= (2\pi)^{-1} \, d_k \times \bigg\{ \begin{matrix} \left[ \ln \left( d_k / 2 \right) - 1 \right] \quad (k = j), \\ \ln \big| \vec{R}_k - \vec{R}_j \big| \qquad (k \neq j), \end{matrix} \label{eq-supp-G-4} \\
\partial G_{j,k} &= \bigg\{ \begin{matrix} -1/2 \qquad \qquad \qquad \qquad \qquad \quad \,\,\, (k = j), \\ (2\pi)^{-1} \, \vartheta (\vec{r}_{k+1} - \vec{R}_j, \vec{r}_{k} - \vec{R}_j) \quad (k \neq j) \, \end{matrix} \nonumber
\end{align}
in terms of the angle $\vartheta (\vec{a}, \vec{b})$ between two general vectors $\vec{a}$ and $\vec{b}$. The discretized version of Eq.~(6) in the main text reads
\begin{equation}
\left( \mathbf{I} + \ell \, \mathbf{D} \right) \cdot \mathbf{T}_{\mathrm{edge}} - \mathbf{T}_{\mathrm{bulk}} = 0, \label{eq-supp-T-3}
\end{equation}
where $\mathbf{I}$ is the $N \times N$ unit matrix, and $\mathbf{D}$ is an $N \times N$ matrix implementing the tangential derivative along the edge:
\begin{equation}
D_{j,k} = \Bigg\{ \begin{matrix} \left( d_k + d_{k-1} \right)^{-1} - \left( d_k + d_{k+1} \right)^{-1} \quad (k = j), \quad \,\,\,\,\, \\ \pm \left( d_j + d_k \right)^{-1} \qquad \qquad \qquad \qquad \,\,\, (k = j \pm 1), \, \\ 0 \qquad \qquad \qquad \qquad \qquad \qquad \qquad (\mathrm{otherwise}). \end{matrix} \label{eq-supp-D}
\end{equation}
Finally, the equations for free, heated, and heat-bath edge segments in Eq.~(7) of the main text readily translate into
\begin{align}
\kappa_{xx} \left( \partial \mathbf{T}_{\mathrm{bulk}} \right)_j + \kappa_{\mathrm{edge}} \left( \mathbf{D} \cdot \mathbf{T}_{\mathrm{edge}} \right)_j  &= \bigg\{ \begin{matrix} 0 \qquad (j \textrm{ is a free segment}), \quad \, \\ p \qquad (j \textrm{ is a heated segment}), \, \end{matrix} \label{eq-supp-T-4} \\
\left( \mathbf{T}_{\mathrm{bulk}} \right)_j &= 0 \qquad \,\,\,\, (j \textrm{ is a heat-bath segment}). \nonumber
\end{align}
Since Eqs.~(\ref{eq-supp-T-2}), (\ref{eq-supp-T-3}), and (\ref{eq-supp-T-4}) contain $3N$ independent linear equations, they provide a unique solution for the $3N$ elements of the vectors $\mathbf{T}_{\mathrm{bulk}}$, $\mathbf{T}_{\mathrm{edge}}$, and $\partial \mathbf{T}_{\mathrm{bulk}}$ in Eq.~(\ref{eq-supp-T-1}). The bulk temperature at a generic point $\vec{r}_0$ inside the edge can then be readily obtained via Eq.~(\ref{eq-supp-G-3}). Note that, since $\ell = \kappa_{\mathrm{edge}} / \lambda$, a sign reversal of $\kappa_{\mathrm{edge}}$ is always accompanied by a sign reversal of $\ell$.

\clearpage

\end{widetext}



\begin{references}

\bibitem{Balents-2010} L. Balents, Nature \textbf{464}, 199 (2010).
\bibitem{Savary-2016} L. Savary and L. Balents, Rep. Prog. Phys. \textbf{80}, 016502 (2016).
\bibitem{Zhou-2017} Y. Zhou, K. Kanoda, and T.-K. Ng, Rev. Mod. Phys. \textbf{89}, 025003 (2017).
\bibitem{Knolle-2019} J. Knolle and R. Moessner, Annu. Rev. Condens. Matter Phys. \textbf{10}, 451 (2019).
\bibitem{Broholm-2020} C. Broholm, R. J. Cava, S. A. Kivelson, D. G. Nocera, M. R. Norman, and T. Senthil, Science \textbf{367}, eaay0668 (2020).
\bibitem{Kitaev-2003} A. Y. Kitaev, Ann. Phys. \textbf{303}, 2 (2003).
\bibitem{Nayak-2008} C. Nayak, S. H. Simon, A. Stern, M. Freedman, and S. Das Sarma, Rev. Mod. Phys. \textbf{80}, 1083 (2008).
\bibitem{Klocke-2024} K. Klocke, Y. Liu, G. B. Hal\'asz, and J. Alicea, arXiv:2411.08093.
\bibitem{Scheie-2024} A. O. Scheie, M. Lee, K. Wang, P. Laurell, E. S. Choi, D. Pajerowski, Q. Zhang, J. Ma, H. D. Zhou, S. Lee, S. M. Thomas, M. O. Ajeesh, P. F. S. Rosa, A. Chen, V. S. Zapf, M. Heyl, C. D. Batista, E. Dagotto, J. E. Moore, and D. A. Tennant, arXiv:2406.17773.
\bibitem{Kalmeyer-1987} V. Kalmeyer and R. B. Laughlin, Phys. Rev. Lett. \textbf{59}, 2095 (1987).
\bibitem{Kalmeyer-1989} V. Kalmeyer and R. B. Laughlin, Phys. Rev. B \textbf{39}, 11879 (1989).
\bibitem{Kitaev-2006} A. Y. Kitaev, Ann. Phys. \textbf{321}, 2 (2006).
\bibitem{Guo-2022} S. Guo, Y. Xu, R. Cheng, J. Zhou, X. Chen, The Innovation, \textbf{3}, 100290 (2022).
\bibitem{Zhang-2024} X.-T. Zhang, Y. H. Gao, and G. Chen, Phys. Rep. \textbf{1070}, 1 (2024).
\bibitem{Kasahara-2018} Y. Kasahara, T. Ohnishi, Y. Mizukami, O. Tanaka, S. Ma, K. Sugii, N. Kurita, H. Tanaka, J. Nasu, Y. Motome, T. Shibauchi, and Y. Matsuda, Nature \textbf{559}, 227 (2018).
\bibitem{Yamashita-2020} M. Yamashita, J. Gouchi, Y. Uwatoko, N. Kurita, and H. Tanaka, Phys. Rev. B \textbf{102}, 220404(R) (2020).
\bibitem{Czajka-2021} P. Czajka, T. Gao, M. Hirschberger, P. Lampen-Kelley, A. Banerjee, J. Yan, D. G. Mandrus, S. E. Nagler, and N. P. Ong, Nat. Phys. \textbf{17}, 915 (2021).
\bibitem{Yokoi-2021} T. Yokoi, S. Ma, Y. Kasahara, S. Kasahara, T. Shibauchi, N. Kurita, H. Tanaka, J. Nasu, Y. Motome, C. Hickey, S. Trebst, and Y. Matsuda, Science \textbf{373}, 568 (2021).
\bibitem{Tanaka-2022} O. Tanaka, Y. Mizukami, R. Harasawa, K. Hashimoto, K. Hwang, N. Kurita, H. Tanaka, S. Fujimoto, Y. Matsuda, E.-G. Moon, and T. Shibauchi, Nat. Phys. \textbf{18}, 429 (2022).
\bibitem{Bruin-2022a} J. A. N. Bruin, R. R. Claus, Y. Matsumoto, N. Kurita, H. Tanaka, and H. Takagi, Nat. Phys. \textbf{18}, 401 (2022).
\bibitem{Lefrancois-2022} \'E. Lefran\c{c}ois, G. Grissonnanche, J. Baglo, P. Lampen-Kelley, J.-Q. Yan, C. Balz, D. Mandrus, S. E. Nagler, S. Kim, Y.-J. Kim, N. Doiron-Leyraud, and L. Taillefer, Phys. Rev. X \textbf{12}, 021025 (2022).
\bibitem{Bruin-2022b} J. A. N. Bruin, R. R. Claus, Y. Matsumoto, J. Nuss, S. Laha, B. V. Lotsch, N. Kurita, H. Tanaka, and H. Takagi, APL Mater. \textbf{10}, 090703 (2022).
\bibitem{Czajka-2023} P. Czajka, T. Gao, M. Hirschberger, P. Lampen-Kelley, A. Banerjee, N. Quirk, D. G. Mandrus, S. E. Nagler, and N. P. Ong, Nat. Mater. \textbf{22}, 36 (2023).
\bibitem{Lefrancois-2023} \'E. Lefran\c{c}ois, J. Baglo, Q. Barth\'elemy, S. Kim, Y.-J. Kim, and L. Taillefer, Phys. Rev. B \textbf{107}, 064408 (2023).
\bibitem{Zhang-2023} H. Zhang, A. F. May, H. Miao, B. C. Sales, D. G. Mandrus, S. E. Nagler, M. A. McGuire, and J. Yan, Phys. Rev. Mater. \textbf{7}, 114403 (2023).
\bibitem{Zhang-2024a} H. Zhang, M. A. McGuire, A. F. May, H.-Y. Chao, Q. Zheng, M. Chi, B. C. Sales, D. G. Mandrus, S. E. Nagler, H. Miao, F. Ye, and J. Yan, Phys. Rev. Mater. \textbf{8}, 014402 (2024).
\bibitem{Imamura-2024}  K. Imamura, S. Suetsugu, Y. Mizukami, Y. Yoshida, K. Hashimoto, K. Ohtsuka, Y. Kasahara, N. Kurita, H. Tanaka, P. Noh, J. Nasu, E.-G. Moon, Y. Matsuda, and T. Shibauchi, Sci. Adv. \textbf{10}, eadk3539 (2024).
\bibitem{Zhang-2024b} H. Zhang, H. Miao, T. Z. Ward, D. G. Mandrus, S. E. Nagler, M. A. McGuire, and J. Yan, Phys. Rev. Lett. \textbf{133}, 206603 (2024).
\bibitem{Ye-2018} M. Ye, G. B. Hal\'asz, L. Savary, and L. Balents, Phys. Rev. Lett. \textbf{121}, 147201 (2018).
\bibitem{Vinkler-Aviv-2018} Y. Vinkler-Aviv and A. Rosch, Phys. Rev. X \textbf{8}, 031032 (2018).
\bibitem{Klocke-2022} K. Klocke, J. E. Moore, J. Alicea, and G. B. Hal\'asz, Phys. Rev. X \textbf{12}, 011034 (2022).
\bibitem{Klocke-2021} K. Klocke, D. Aasen, R. S. K. Mong, E. A. Demler, and J. Alicea, Phys. Rev. Lett. \textbf{126}, 177204 (2021).
\bibitem{Wei-2021} Z. Wei, V. F. Mitrovi\'c, and D. E. Feldman, Phys. Rev. Lett. \textbf{127}, 167204 (2021).
\bibitem{Wei-2023} Z. Wei, N. Batra, V. F. Mitrovi\'c, and D. E. Feldman, Phys. Rev. B \textbf{107}, 104406 (2023).
\bibitem{Zhang-2025} H. Zhang \emph{et al.}, in preparation.
\bibitem{Strohm-2005} C. Strohm, G. L. J. A. Rikken, and P. Wyder, Phys. Rev. Lett. \textbf{95}, 155901 (2005).
\bibitem{Li-2020} X. Li, B. Fauqu\'e, Z. Zhu, and K. Behnia, Phys. Rev. Lett. \textbf{124}, 105901 (2020).
\bibitem{Grissonnanche-2020} G. Grissonnanche, S. Th\'eriault, A. Gourgout, M.-E. Boulanger, \'E. Lefran\c{c}ois, A. Ataei, F. Lalibert\'e, M. Dion, J.-S. Zhou, S. Pyon, T. Takayama, H. Takagi, N. Doiron-Leyraud, and L. Taillefer, Nat. Phys. \textbf{16}, 1108 (2020).
\bibitem{Boulanger-2020} M.-E. Boulanger, G. Grissonnanche, S. Badoux, A. Allaire, \'E. Lefran\c{c}ois, A. Legros, A. Gourgout, M. Dion, C. H. Wang, X. H. Chen, R. Liang, W. N. Hardy, D. A. Bonn, and L. Taillefer, Nat. Commun. \textbf{11}, 5325 (2020).
\bibitem{Onose-2010} Y. Onose, T. Ideue, H. Katsura, Y. Shiomi, N. Nagaosa, and Y. Tokura, Science \textbf{329}, 297 (2010).
\bibitem{Hirschberger-2015} M. Hirschberger, R. Chisnell, Y. S. Lee, and N. P. Ong, Phys. Rev. Lett. \textbf{115}, 106603 (2015).
\bibitem{SM} Supplemental Material.
\bibitem{Katsura-2010} H. Katsura, N. Nagaosa, and P. A. Lee, Phys. Rev. Lett. \textbf{104}, 066403 (2010).
\bibitem{Matsumoto-2011} R. Matsumoto and S. Murakami, Phys. Rev. Lett. \textbf{106}, 197202 (2011).
\bibitem{Habel-2024} J. Habel, A. Mook, J. Willsher, and J. Knolle, Phys. Rev. B \textbf{109}, 024441 (2024).

\end{references}
\end{document}